\documentclass[aps,prl,twocolumn,superscriptaddress]{revtex4}
\usepackage{graphicx}
\usepackage{amsmath}
\usepackage{amsfonts}
\usepackage{color}
\usepackage{bm}% bold math
\newcommand{\ket}[1]{\ensuremath{\left|{#1}\right\rangle}}
\newcommand{\bra}[1]{\ensuremath{\left\langle{#1}\right |}}

\newcommand{\oper}[1]{\mathbf{\mathsf{#1}}}

\newcommand{\beq}{\begin{equation}}
\newcommand{\eeq}{  \end{equation}}
\newcommand{\bea}{\begin{eqnarray}}
\newcommand{\eea}{  \end{eqnarray}}
\newcommand{\bit}{\begin{itemize}}
\newcommand{\eit}{  \end{itemize}}

\begin{document}

%\preprint{}

\title{Entropic Entanglement Criteria for Continuous Variables}

%--------------------------------------
\author{S. P. Walborn}
\affiliation{Instituto de F\'{\i}sica, Universidade Federal do Rio
de Janeiro, Caixa Postal 68528, Rio de Janeiro, RJ 21941-972,
Brazil}
%--------------------------------------
\author{B. G. Taketani}
\affiliation{Instituto de F\'{\i}sica, Universidade Federal do Rio
de Janeiro, Caixa Postal 68528, Rio de Janeiro, RJ 21941-972,
Brazil}
%--------------------------------------
\author{A. Salles}
\affiliation{Instituto de F\'{\i}sica, Universidade Federal do Rio
de Janeiro, Caixa Postal 68528, Rio de Janeiro, RJ 21941-972,
Brazil}
%---------------------------------------
\author{F. Toscano}
 \affiliation{Instituto de F\'{\i}sica, Universidade Federal do Rio
              de Janeiro, Caixa Postal 68528, Rio de Janeiro, RJ 21941-972,
              Brazil}
%--------------------------------------
\author{R. L. de Matos Filho}
\affiliation{Instituto de F\'{\i}sica, Universidade Federal do Rio
de Janeiro, Caixa Postal 68528, Rio de Janeiro, RJ 21941-972,
Brazil}

\date{\today}

\begin{abstract}
We derive several entanglement criteria for bipartite continuous variable quantum systems based on the Shannon entropy.  These criteria are more sensitive than those involving only second-order moments, and are equivalent to well-known variance product tests in the case of Gaussian states.   Furthermore, they involve only a pair of quadrature measurements, { and will thus prove extremely useful %convenient
in }the experimental identification of entanglement.  
  \end{abstract}

%\keywords{}

\maketitle
\par
Quantum entanglement is the property that differentiates quantum mechanical systems from classical ones.  As such, the detection and characterization of quantum entanglement is one of the prominent goals in Quantum Information.  In the discrete variable case, many detection schemes for entanglement have been proposed (see \cite{guhne09} for review).  In the continuous variable (CV) case, detection of entanglement is more challenging due to the { complicated} Hilbert space structure, and many tests only detect entanglement that appears in the second-order moments \cite{simon00,duan00,mancini02,giovannetti03,guhne04,hyllus06}, which is completely adequate for the case of Gaussian states.  However, non-gaussian states and processes have been shown to not just enhance certain quantum information protocols such as teleportation \cite{opatrny,dellanno}, but in fact be necessary for certain tasks, such as universal quantum computing \cite{braunstein99,Bartlett02} and entanglement distillation \cite{dong08,hage08}. Towards the detection of CV entanglement in general, Shchukin and Vogel have derived an infinite hierarchy of conditions for positive partial transpose involving higher-order moments \cite{shchukin05}.  Although powerful, these conditions may not always be experimentally convenient \cite{shchukin05b}.    
\par
Here we derive several entropic entanglement criteria for CVs.  In contrast to previous work based on quantum-mechanical generalizations of entropy functions \cite{barnett89,horodecki96a,cerf97,vollbrecht02}, our criteria involve the Shannon entropy of probability distributions of a pair of complementary quadrature measurements.  We will show that these conditions detect entanglement in many states that any second-order test will not.  A first set of inequalities is most sensitive, but is valid only for pure states.  Inspired by previous work in discrete variables \cite{giovannetti04},  we use the entropic uncertainty relations for complementary CV observables \cite{bialynicki75} to derive a second set of inequalities. These have the distinct advantage that they can be extended to include mixed bipartite CV states.  These inequalities are more sensitive than the usual criteria based on second-order moments \cite{simon00,duan00,mancini02,giovannetti03}, and are equivalent to a well-known variance product criteria \cite{mancini02} in the case of bipartite Gaussian states. { At the same time they are no more experimentally demanding than the widely adopted tests~\cite{duan00,mancini02,giovannetti03}.}      
 
\par
As in other CV inseparability criteria \cite{duan00,mancini02,giovannetti03}, we consider the global operators 
\begin{subequations}
\label{eq:r}
\begin{equation}
\oper{r}_\pm = \oper{r}_1 \pm  \oper{r}_2
\end{equation}
\begin{equation}
\oper{s}_\pm = \oper{s}_1 \pm \oper{s}_2,
\end{equation}
\end{subequations}
 where $\oper{r}_j=\cos\theta_j \oper{x_j} + \sin\theta_j \oper{p}_j$, $\oper{s}_j =\cos\theta_j \oper{p_j} - \sin\theta_j \oper{x}_j$ and $\oper{x}_j$ and $\oper{p}_j$ are the usual canonical variables satisfying $[\oper{x}_j,\oper{p}_i]=i \delta_{ij}$, and $j,i=1,2$ refers to each subsystem of a bipartite state.  Note also that  the operators $\oper{r}_j$ and $\oper{s}_j$  satisfy $[\oper{r}_j,\oper{s}_i]=i \delta_{ij}$.  The entropy associated to a measurement of $\oper{r}$ is given by the Shannon entropy
\begin{equation}
H[R] = - \int dr R(r) \ln R(r), 
\end{equation}
where $R(r)$ is the probability distribution associated to the measurement of $\oper{r}$, and similarly for $H[S]$.  
\par
We will derive inseparability criteria of the form
\begin{equation}
H[R_\pm] + H[S_\mp] \geq c,  
\label{eq:sepcritgen}
\end{equation}
where $R_\pm$ and $S_\mp$ are the probability distributions associated to measurement of $\oper{r_\pm}$ and $\oper{s_\mp}$, respectively, and  $c>0$ is a real constant.  Any separable state obeys inequality \eqref{eq:sepcritgen}, while entangled states may not.  For example, the left side of Eq. \eqref{eq:sepcritgen} vanishes for the common eigenstates of $\oper{r}_-$ and $\oper{s}_+$ { or $r_+$ and $s_-$, which correspond EPR-like states
(note that $[r_{+},s_{-}]=[r_{-},s_{+}]=0$)}.   
\par  
 Let us first derive an inequality for the case of pure states.   We will then extend our results to include mixed states.  A separable pure state can be written in the form $\ket{\psi_{1}} \otimes  \ket{\psi_{2}}$, and has a corresponding wave function $\Psi(r_1,r_2) = \psi_1(r_1) \psi_2(r_2)$.   Using Eq. \eqref{eq:r} and changing variables gives
\begin{equation}
\Psi(r_+,r_-) = \frac{1}{\sqrt{2}}\psi_1\left(\frac{r_+ +r_-}{2}\right)\psi_2\left(\frac{r_+  - r_-}{2}\right).  
\end{equation}
The probability distribution associated to the measurement of  $\oper{r}_\pm$ is given by
\begin{align}
R_\pm & = \frac{1}{2} \int dr_\mp R_1\left(\frac{r_+ +r_-}{2}\right)R_2\left(\frac{r_+  - r_-}{2}\right),  \nonumber \\ 
& =  \int dr R_1(r)R_2(\mp r  \pm r_\pm)=R_1*R_2^{(\pm)},     
\end{align}
{where $R_j(r)=|\psi_j(r)|^2$,  the symbol ``$\ast$" denotes convolution and
$R_2^{(+)}\equiv R_2(r)$, $R_2^{(-)}\equiv R_2(-r)$.
%Equivalently, we can show that the probability distribution 
%associated with the measurement of $\oper{s}_{\pm}$ is given by
%$S_\pm =S_1 \ast S_2^{(\pm)}$ (where $S_2^{(+)}\equiv S_2(r)$ and  $S_2^{(-)}\equiv S_2(-r)$).
%for the entropy associated with $\oper{r}_\pm$ measurements
%$H[R_\pm]   =H[R_1 \ast R_2^{(\pm)}] =H[R_1 \ast R_2]$. 
%and $H[S_\pm]   =H[S_1 \ast S_2^{(\pm)}] =H[S_1 \ast S_2]$. }{\bf coloquei assim embora 
%nao consigo provar!} 
Using the entropy power inequality~\cite{shannon,cover} 
\begin{equation}
\exp(2 H[A \ast B]) \geq \exp(2H[A]) + \exp(2H[B]),
\label{eq:entpow}
\end{equation}
and also the fact that the Shannon entropy is invariant under reflections~\cite{shannon}, we have
\begin{equation}
H[R_\pm]   \geq  \frac{1}{2} \ln \left \{ \exp(2 H[R_1])   +  \exp(2 H[R_2]) \right \}. 
\label{eq:HP}
\end{equation}
} %end blue
We arrive at an equivalent inequality for $H[S_\mp]$:
\begin{equation}
H[S_\mp]   \geq  \frac{1}{2} \ln \left \{ \exp(2 H[S_1])   +  \exp(2 H[S_2]) \right \}. 
\label{eq:HQ}
\end{equation}
Combining \eqref{eq:HP} and \eqref{eq:HQ}, we have
\begin{equation}
  \label{eq:sum}
H[R_\pm]  + H[S_\mp]  \geq    \frac{1}{2} \ln \left \{ \sum\limits_{i,j=1,2} e^{(2 H[R_i]+ 2 H[S_j])}  \right \}. 
\end{equation}
 Eq. \eqref{eq:sum} gives two inequalities that are satisfied by separable pure states.  Violation of either inequality \eqref{eq:sum} is then a sufficient condition for entanglement.   
 \par
 We will now show that it is possible to arrive at a weaker pair of inequalities, and then extend them to include mixed states.    
Using the entropic uncertainty relation for continuous variables ($j=1,2$) \cite{bialynicki75}, 
\begin{equation}
H[R_j] + H[S_j] \geq \ln \pi e, 
\label{eq:uncent}
\end{equation}
gives $\exp(2H[R_j] + 2H[S_j]) \geq (\pi e)^2$, which leads to
\begin{align}
H[R_\pm]  + H[S_\mp]  \geq  &  \frac{1}{2} \ln \left \{ 2 (\pi e)^2 + \sum\limits_{i \neq j} e^{(2 H[R_i]+ 2 H[S_j])} \right \}. \nonumber \\
\label{eq:sep2}
\end{align}
Using relation \eqref{eq:uncent} again, one obtains
\begin{align}
H[R_\pm]  + H[S_\mp]  \geq  &  \frac{1}{2} \ln \left \{ 2 (\pi e)^2 + \right. 
 \nonumber \\
& \left. 2 (\pi e)^2 \cosh \left ( 2H[S_2] - 2H[S_1]  \right )   \right \}. \nonumber \\
\label{eq:sepcrit}
\end{align}
Since the hyperbolic cosine is lower-bounded by $1$, we have
\begin{equation}
H[R_\pm]  + H[S_\mp]   \geq  \ln 2 \pi e. 
\label{eq:sepind}
\end{equation}
The term on the right side is a state-independent lower bound for $H[R_\pm]  + H[S_\mp]$ of any separable pure state. 
\par
To extend the inequalities \eqref{eq:sepind} to include mixed states, we use the fact that any bipartite separable state $\rho$ can be decomposed into a convex sum of pure states 
\begin{equation}
\rho = \sum_{k} \lambda_k \ket{\psi_{1k}}\bra{\psi_{1k}} \otimes  \ket{\psi_{2k}}\bra{\psi_{2k}},   
\label{eq:rho}
\end{equation}
where $\lambda_k \geq 0$ and $\sum_k \lambda_k=1$. The probability distributions associated to a measurement $r_{\pm}$ is 
\begin{equation}
R_\pm = \sum_{k} \lambda_k R_{k\pm}.    
\label{eq:prob}
\end{equation}
Here $R_{k\pm}$ is the probability to detect $r_\pm$ for each pure state in the decomposition \eqref{eq:rho}. 
The concavity of the  Shannon entropy \cite{cover}  gives
\begin{equation}
H[R_\pm] \geq \sum\limits_k \lambda_k H[R_{k\pm}],   
\end{equation}
 and likewise for $H[S_\pm]$.
As $H[R_{k\pm}]=H[R_{k1}\ast R_{k2}]$, inequality \eqref{eq:entpow} gives
\begin{equation}
H[R_\pm] \geq \sum\limits_k \frac{\lambda_k}{2} \ln \left ( e^{2H[R_{k1}]} + e^{2H[R_{k2}]} \right).   
\end{equation}
A similar condition holds for $H[S_\pm]$. 
Summing these two inequalities, and using the fact that the left side of \eqref{eq:sum} is lower-bounded by $\ln 2 \pi e$, gives
\begin{align}
H[R_\pm]  + H[S_\mp]  \geq  &  \sum\limits_k \frac{\lambda_k}{2} \ln \left \{ \sum\limits_{i,j=1,2} e^{(2 H[R_{ki}]+ 2 H[S_{kj}])} \right \}. \nonumber  \\
& \geq \sum \limits_k \lambda_k \ln(2 \pi e) = \ln 2 \pi e,
  \label{eq:summix}
\end{align}
which is identical to \eqref{eq:sepind}. Thus, inequalities \eqref{eq:sepind} are satisfied by both pure and mixed separable states.   We note also that one can take the supremum of the  first inequality \eqref{eq:summix} over all possible decompositions of the mixed state $\rho$ to arrive at a stronger inequality.  However, this is not suitable for experimental purposes.   

We also note that inequalities \eqref{eq:sepind}  can also be obtained using the positive partial transpose criterion \cite{simon00} as we will now show. 
{ First, we note that the marginal distributions under partial transposition are:
 \begin{equation}
 \label{eq:PPTcondition}
 \tilde{R}_{\pm}=R_{\pm}\;\; \mbox{and}\;\;   \tilde{S}_{\pm}= S_{\mp}\;.
 \end{equation}  
Noting that $[\oper{r}_{\pm},\oper{s}_{\pm}]=2 i$, the relations in~\eqref{eq:PPTcondition}
imply that any separable state must verify the uncertainty relation 
$H[\tilde{R}_{\pm}] + H[\tilde{S}_{\pm}]\geq \ln 2 \pi e$, which leads directly to inequalities~\eqref{eq:sepind}. 
Furthermore, we see in~\eqref{eq:PPTcondition} that the partial transposition %flip the role
interchanges the variables $s_{+}$ and $s_{-}$, %so this 
which is the key to obtain 
entanglement criteria based on uncertainty inequalities that can %steam 
arise from the  
noncomutativity of $\oper{r}_{\pm}$ and $\oper{s}_{\pm}$.}
%{\bf enfatizei mais este resultado por um lado para reforçar a validade da Eq. (13) cuja demostracao tem uma parte que nao esta muito clara e
%tambem pois acho que e' um resultado teorico que deveriamos enfatizar nas conclusoes.}
       
\par 
An upper bound to the left side of \eqref{eq:sepind} can be obtained by considering that the Shannon entropy of a continuous variable with variance $\sigma^2$ is maximized when the probability distribution is Gaussian, for which $H_{\mathrm{Gauss}}=\ln \sqrt{2 \pi e \sigma^2}$ \cite{shannon,cover}.  Then, 
\begin{equation}
\ln {2 \pi e \sigma_\pm \delta_\mp} \geq H[R_\pm]  + H[S_\mp]   \geq \ln 2 \pi e,   
\label{eq:bounds}
\end{equation}
where $\sigma_{\pm}^2$ and $\delta_\pm^2$ are the variances of $R_\pm$ and $S_\pm$, respectively.  
This upper limit is reached for Gaussian states, in which case we recover the Mancini-Giovannetti-Vitali-Tombesi (MGVT) product inequality \cite{mancini02}
\begin{equation}
\sigma_\pm \delta_\mp \geq 1.  
\label{eq:mancini}
\end{equation}
The left side of the double inequality \eqref{eq:bounds} proves that the conditions \eqref{eq:sepind} are more sensitive than the variance product criteria, with equivalence in the case of Gaussian states.  This is in accord with the extremality of entangled Gaussian states \cite{wolf06}. 
\par
\par
In the definition of the operators \eqref{eq:r} we included the parameter $\theta_j$ to account for local rotations of the quadrature measurements.  
To successfully employ the entropic criteria, it is necessary to find suitable {rotated} quadrature operators, parametrized by angles $\theta_1$ and $\theta_2$.   In addition, one can optimize over local squeezing parameters, which can be included a posteriori \cite{hyllus06}.   
{ So, in order to complete the analysis of the effect of real linear canonical transformations over the single mode quadrature 
operators (that form the real symplectic group $Sp(2,\mathbb{R})$)} let us consider now the 
effect of local squeezing, which can be accounted for by redefining the 
{{ rotated} operators as
\begin{subequations}
\label{eq:ra}
\begin{equation}
\oper{r}_\pm^\prime = a_1\oper{r}_1 \pm  a_2 \oper{r}_2
\end{equation}
\begin{equation}
\oper{s}_\pm^\prime =\frac{1}{a_1} \oper{s}_1 \pm \frac{1}{a_2}\oper{s}_2.
\end{equation}
\end{subequations}
 Substituting these operators, inequality  \eqref{eq:sum} becomes
 \begin{equation}
  \label{eq:suma}
H[R_\pm^\prime]  + H[S_\mp^\prime]  \geq    \frac{1}{2} \ln \left \{ \sum\limits_{i,j=1,2} e^{(2 H[R_i]+ 2 H[S_j])+2\ln\frac{a_i}{a_j}}  \right \},    
\end{equation} 
and applying the entropic uncertainty relation \eqref{eq:uncent} results in 
 \begin{equation}
H[R_\pm^\prime]  + H[S_\mp^\prime]  \geq  \ln(2 \pi e),   
\label{eq:sepinda}
\end{equation}
where $R_\pm^\prime$ and $S_\pm^\prime$ are the probability distributions for measurements of the operators \eqref{eq:ra}.  It is straightforward to show that both entropy inequalities \eqref{eq:suma} and \eqref{eq:sepinda} reduce to \eqref{eq:sum} and \eqref{eq:sepind} when $a_1=a_2$, which demonstrates that these inequalities are invariant to equal amounts of local squeezing. 
\par
\textit{Examples.}
Inequalities \eqref{eq:sum} and \eqref{eq:sepind} are more sensitive than criteria involving sums or products of variances.  We will now illustrate this point with several examples of non-Gaussian states.  
Let us first consider a non-gaussian wave function of the form  
\begin{equation}
\eta(r_1,r_2) = \frac{(r_1+r_2)}{\sqrt{\pi \sigma_- \sigma_+^3}}e^{-(r_1+r_2)^2/4\sigma_+^2}e^{-(r_1-r_2)^2/4\sigma_-^2},
\label{eq:eta}
\end{equation} 
This state is non-separable for all values of $\sigma_\pm$.  For this state both the Simon PPT criteria \cite{simon00}, which is { a necessary and sufficient condition to detect
entanglement in Gaussian states}, and the MGVT criteria with $\theta_1=\theta_2=0$, detect entanglement provided 
{$\sigma_-/\sigma_{+} > \sqrt{3} \approx 1.732$ or
$\sigma_-/\sigma_{+} < 1/\sqrt{3} \approx 0.577$}.
%$\sigma_\pm/\sigma_\mp < 1/\sqrt{3} \approx 0.577$. 
With $\theta_1=\theta_2=0$, the inseparability criteria \eqref{eq:sepind} gives
$H[R_\pm] + H[S_\mp] = \ln \left(4 \pi e^{\gamma} {\sigma_\mp}/{\sigma_\pm} \right)$,  where $\gamma = 0.577\dots$ is Euler's constant.  Entanglement is detected provided 
%$\sigma_\mp/\sigma_\pm < e^{(1-\gamma)}/2 \approx0.763$.  
{ $\sigma_{-}/\sigma_{+} < e^{(1-\gamma)}/2 \approx0.763$
or $\sigma_{-}/\sigma_{+} > 2/e^{(1-\gamma)} \approx 1.310$.}  
Thus, the entropy criterion \eqref{eq:sepind} is more sensitive than the Simon and MGVT conditions.  Numerical results show that the ranges in which the pair of inequalities  \eqref{eq:sum} detect entanglement overlap, indicating that they always detect entanglement in the state \eqref{eq:eta}.    
A pictoral representation of these results is shown in FIG. \ref{fig:hg}.  
  %%%%%%%%%%%%%%%%%%%
  \begin{figure}
  \begin{center}
  \includegraphics[width=7cm]{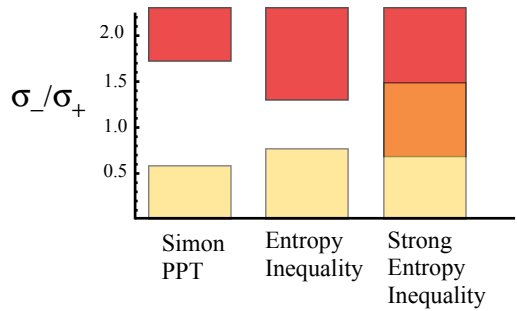}
  \caption{\label{fig:hg} (Color online) Pictoral representation of three inseparability criteria for the pure state \eqref{eq:eta} as a function of $\sigma_-/\sigma_+$. 
{ The criteria called ``Entropy inequality''  and ``Strong Entropy inequality'' correspond 
respectively to  Eqs.(\ref{eq:sepind}) and (\ref{eq:sum}).   
The dark and light  grey regions correspond to the intervals of  $\sigma_-/\sigma_+$ 
where each  criteria detect entanglement and blank regions where they do not.}  
Both criteria presented here are stronger than the Simon PPT condition \cite{simon00}.}
  \end{center}
  \end{figure}
  %%%%%%%%%%%%%%%%%%%
  \par
  We tested these criteria numerically for a number of combinations of low-order Fock states. 
 A particularly interesting example is the state $\ket{\phi}=\ket{0,0}/\sqrt{2}+\ket{2,0}/2 + \ket{0,2}/2$, which is undetectable with any second-order criterion \cite{rodo08}.  Entanglement is detectable using the stronger entropic inequality \eqref{eq:sum} with $\theta_1=-\pi/4$ and $\theta_2=\pi/4$. 
We also tested ``N00N" states of the form $(\ket{N,0}+\ket{0,N})/\sqrt{2}$, where $\ket{N}$ is an $N$-photon Fock state. None of these states is detected by the Simon PPT criteria \cite{simon00} nor inequalities \eqref{eq:sepind}. The strong criteria \eqref{eq:sum} detects entanglement up to $N=5$  with $\theta_1=\theta_2=0$ for all values of $N$ except $N=2$, for which $\theta_1=0$ and $\theta_2=\pi/2$ were used.     
\par
To further test these criteria we generated uniform random pure states \cite{zyczkowski01} of the form $\ket{\psi}=\sum_{n,m=0}^D C_{nm}\ket{n,m}$, and tested inequalities \eqref{eq:sum}, \eqref{eq:sepind} and also the MGVT criteria, while varying angles $\theta_1$ and $\theta_2$ in intervals of $\pi/4$.  Results are summarized in table \ref{tab:1}. The strong inequality \eqref{eq:sum} {detected more than $ 62 \%$ of the states up to $D=7$}.
%%%%%%%%%%%%%%%%%%%%%%%%
\begin{table}
 \begin{ruledtabular}
 \begin{tabular}{ccccc}
\# states & $D$ & $n_{strong}$ & $n_{weak}$ & $n_{MGVT}$ \\
\hline
6000 & 2 & 74.4\% & 17.3 \% & 9.9 \% \\
1600 & 3 & 86.3\% & 0.5 \% & 0.2 \% \\
800 & 4 & 84.9 \% & 0\% & 0\% \\
720 & 5 & 81.0 \% & 0\% & 0\% \\
120 & 7 & 62.5\% & 0\% & 0\% \\
\end{tabular}
 \end{ruledtabular}
 \caption{\label{tab:1}  Results for random states $\ket{\psi}$ (see text). $n_{strong}$, $n_{weak}$ and $n_{MGVT}$ are the percentage of states detected by inequalities \eqref{eq:sum}, \eqref{eq:sepind} and \eqref{eq:mancini}, respectively.  In all cases, the angles $\theta_1$ and $\theta_2$ were scanned in intervals of $\pi/4$.}
\end{table}
\par
Criteria \eqref{eq:sepind} also applies to mixed states.   Let us consider a dephased cat state characterized by the parameter $0 \leq p \leq 1$, given by
\begin{align}
\rho =  & N(\alpha) \left \{ \ket{\alpha,\alpha}\bra{\alpha,\alpha} + \ket{-\alpha,-\alpha}\bra{-\alpha,-\alpha}  \right. \nonumber \\
& \left. -  (1-p) (\ket{\alpha,\alpha}\bra{-\alpha,-\alpha} +\ket{-\alpha,-\alpha}\bra{\alpha,\alpha} ) \right \},
\label{eq:cat}
\end{align}
where $N(\alpha)$ is a normalization constant.  This state is separable only when $p=1$, and is undetected by any second order criteria for any value of $p$. The entanglement criteria \eqref{eq:sepind} is shown in FIG. \ref{fig:cats} as a function of $\alpha$ and $p$ for real $\alpha$.  Using $\theta_1=\theta_2=0$, the sum $H[R_-] + H[S_+]$ is less than $\ln 2 \pi e$, and thus entanglement is detected, for a large range of $\alpha$ and $p$.      
%%%%%%%%%%%%%%%%%%%
  \begin{figure}
  \begin{center}
  \includegraphics[width=7cm]{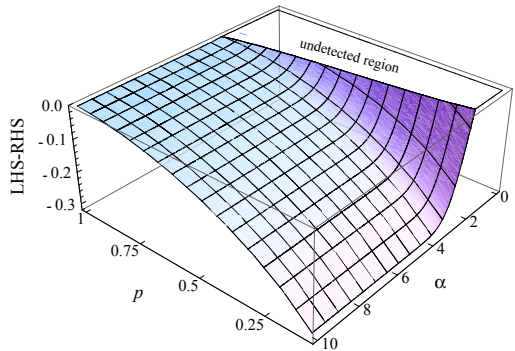}
  \caption{\label{fig:cats} (Color online) Violation of entanglement criteria \eqref{eq:sepind} for the dephased cat state \eqref{eq:cat}.  The vertical axis is the difference of the left-hand side (LHS) and right-hand side (RHS) of \eqref{eq:sepind}.  See text for details.}  
  \end{center}
  \end{figure}
  %%%%%%%%%%%%%%%%%%%
\par
Let us briefly discuss the application of these entropic criteria in an experimental setting.  
{ For fixed values $\theta_1$ and $\theta_2$  of the local rotations} the Shannon entropies  $H[R_\pm]$ and $H[S_\mp]$ can be calculated using the marginal probability distributions $R_\pm$ and $S_\mp$. 
%in the same way as the second-order moments. 
These can be determined directly via measurement of $\oper{r}_\pm$ and $\oper{s}_\mp$, or calculated from the joint probability distributions $R(r_1,r_2)$ and $S(s_1,s_2)$.  
{We %note 
stress that the sole determination of these probability distributions does not allow one to 
calculate arbitrary moments involving products of the $\oper{r}_j$ and $\oper{s}_j$ operators.  
Hence, these measurements alone are not enough to determine all the second-order 
moments required to evaluate even the second-order criterion of Simon~\cite{simon00}.
%to evaluate for example the second-order inequality in  \cite{simon00}
The higher-order criteria in~\cite{shchukin05} requires even more involved measurement schemes~\cite{shchukin05b}.
%neither to apply the criteria in \cite{shchukin05}. 
In summary, the evaluation of our entropic criteria requires the same experimental resources as those required to evaluate the commonly employed second-order inequalities in~\cite{duan00,mancini02,giovannetti03}, while providing a more sensitive entanglement test. We thus expect that the inseparability tests presented here will be of great use in experimental settings.}

%, the experimental resources required to test our criteria are the same as those required to evaluate the second-order inequalities in \cite{duan00,mancini02,giovannetti03}, 

%Furthermore, the experimental resources required to apply our inequalities are the same to those required to evaluate the second-order inequalities in \cite{duan00,mancini02,giovannetti03} (as their are based on the variances of the marginal probability distributions $H[R_\pm]$ and $H[S_\mp]$).}  
%In this regard, we expect that the inseparability criteria above should be useful for experimental identification of entanglement.
\par
%In summary, we have presented several entanglement criteria for bipartite continuous variable states involving the Shannon entropy {(comprised in Eqs. (\ref{eq:sum}) and (\ref{eq:sepind}))}, and shown that these criteria are more sensitive than those involving only second-order moments.  We have illustrated this point with several examples.  In the particular case of gaussian states, these criteria reduce to well-known variance product criteria \cite{mancini02,giovannetti03}. 
%{ We have also shown the relation of our criterion in Eq.(\ref{eq:sepind})
%with the positive partial transpose criterion.}
%Requiring no further resources than the typically used, less sensitive second-order criteria, we expect our novel entropic criteria to be widely adopted in the experimental characterization of continuous variable entanglement.

%We expect that these novel entropic entanglement criteria will be of great use 
%allow for the further characterization of continuous variable entanglement.   

\begin{acknowledgements}
Financial support was provided by Brazilian agencies CNPq and
FAPERJ.
\end{acknowledgements}

%%%%%%%%%%%%%%%%%%%%%BIBLIOGRAPHY%%%%%%%%%%%%%%%%%%%%%%%%%%%%%%%
%\bibliographystyle{apsrev}
%\bibliography{EntropyCVEnt}

\begin{thebibliography}{26}
\expandafter\ifx\csname natexlab\endcsname\relax\def\natexlab#1{#1}\fi
\expandafter\ifx\csname bibnamefont\endcsname\relax
  \def\bibnamefont#1{#1}\fi
\expandafter\ifx\csname bibfnamefont\endcsname\relax
  \def\bibfnamefont#1{#1}\fi
\expandafter\ifx\csname citenamefont\endcsname\relax
  \def\citenamefont#1{#1}\fi
\expandafter\ifx\csname url\endcsname\relax
  \def\url#1{\texttt{#1}}\fi
\expandafter\ifx\csname urlprefix\endcsname\relax\def\urlprefix{URL }\fi
\providecommand{\bibinfo}[2]{#2}
\providecommand{\eprint}[2][]{\url{#2}}

\bibitem[{\citenamefont{G\"uhne and T\'oth}(2009)}]{guhne09}
\bibinfo{author}{\bibfnamefont{O.}~\bibnamefont{G\"uhne}} \bibnamefont{and}
  \bibinfo{author}{\bibfnamefont{G.}~\bibnamefont{T\'oth}},
  \bibinfo{journal}{Phys. Rep.} \textbf{\bibinfo{volume}{474}},
  \bibinfo{pages}{1} (\bibinfo{year}{2009}).

\bibitem[{\citenamefont{Simon}(2000)}]{simon00}
\bibinfo{author}{\bibfnamefont{R.}~\bibnamefont{Simon}},
  \bibinfo{journal}{Phys. Rev. Lett.} \textbf{\bibinfo{volume}{84}},
  \bibinfo{pages}{2726} (\bibinfo{year}{2000}).

\bibitem[{\citenamefont{Duan et~al.}(2000)\citenamefont{Duan, Giedke, Cirac,
  and Zoller}}]{duan00}
\bibinfo{author}{\bibfnamefont{L.-M.} \bibnamefont{Duan}},
  \bibinfo{author}{\bibfnamefont{G.}~\bibnamefont{Giedke}},
  \bibinfo{author}{\bibfnamefont{J.~I.} \bibnamefont{Cirac}}, \bibnamefont{and}
  \bibinfo{author}{\bibfnamefont{P.}~\bibnamefont{Zoller}},
  \bibinfo{journal}{Phys. Rev. Lett.} \textbf{\bibinfo{volume}{84}},
  \bibinfo{pages}{2722} (\bibinfo{year}{2000}).

\bibitem[{\citenamefont{Mancini et~al.}(2002)\citenamefont{Mancini,
  Giovannetti, Vitali, and Tombesi}}]{mancini02}
\bibinfo{author}{\bibfnamefont{S.}~\bibnamefont{Mancini}},
  \bibinfo{author}{\bibfnamefont{V.}~\bibnamefont{Giovannetti}},
  \bibinfo{author}{\bibfnamefont{D.}~\bibnamefont{Vitali}}, \bibnamefont{and}
  \bibinfo{author}{\bibfnamefont{P.}~\bibnamefont{Tombesi}},
  \bibinfo{journal}{Physical Review Letters} \textbf{\bibinfo{volume}{88}},
  \bibinfo{eid}{120401} (\bibinfo{year}{2002}).

\bibitem[{\citenamefont{Giovannetti et~al.}(2003)\citenamefont{Giovannetti,
  Mancini, Vitali, and Tombesi}}]{giovannetti03}
\bibinfo{author}{\bibfnamefont{V.}~\bibnamefont{Giovannetti}},
  \bibinfo{author}{\bibfnamefont{S.}~\bibnamefont{Mancini}},
  \bibinfo{author}{\bibfnamefont{D.}~\bibnamefont{Vitali}}, \bibnamefont{and}
  \bibinfo{author}{\bibfnamefont{P.}~\bibnamefont{Tombesi}},
  \bibinfo{journal}{Phys. Rev. A} \textbf{\bibinfo{volume}{67}},
  \bibinfo{pages}{022320} (\bibinfo{year}{2003}).

\bibitem[{\citenamefont{G\"uhne}(2004)}]{guhne04}
\bibinfo{author}{\bibfnamefont{O.}~\bibnamefont{G\"uhne}},
  \bibinfo{journal}{Phys. Rev. Lett.} \textbf{\bibinfo{volume}{92}},
  \bibinfo{pages}{117903} (\bibinfo{year}{2004}).

\bibitem[{\citenamefont{Hyllus and Eisert}(2006)}]{hyllus06}
\bibinfo{author}{\bibfnamefont{P.}~\bibnamefont{Hyllus}} \bibnamefont{and}
  \bibinfo{author}{\bibfnamefont{J.}~\bibnamefont{Eisert}},
  \bibinfo{journal}{New J. Phys.} \textbf{\bibinfo{volume}{8}},
  \bibinfo{pages}{51} (\bibinfo{year}{2006}).

\bibitem[{\citenamefont{Opatrn\'y et~al.}(2000)\citenamefont{Opatrn\'y,
  Kurizki, and Welsch}}]{opatrny}
\bibinfo{author}{\bibfnamefont{T.}~\bibnamefont{Opatrn\'y}},
  \bibinfo{author}{\bibfnamefont{G.}~\bibnamefont{Kurizki}}, \bibnamefont{and}
  \bibinfo{author}{\bibfnamefont{D.-G.} \bibnamefont{Welsch}},
  \bibinfo{journal}{Phys. Rev. A} \textbf{\bibinfo{volume}{61}},
  \bibinfo{pages}{032302} (\bibinfo{year}{2000}).

\bibitem[{\citenamefont{Dell'Anno et~al.}(2007)\citenamefont{Dell'Anno, Siena,
  Albano, and Illuminati}}]{dellanno}
\bibinfo{author}{\bibfnamefont{F.}~\bibnamefont{Dell'Anno}},
  \bibinfo{author}{\bibfnamefont{S.~D.} \bibnamefont{Siena}},
  \bibinfo{author}{\bibfnamefont{L.}~\bibnamefont{Albano}}, \bibnamefont{and}
  \bibinfo{author}{\bibfnamefont{F.}~\bibnamefont{Illuminati}},
  \bibinfo{journal}{Physical Review A} \textbf{\bibinfo{volume}{76}},
  \bibinfo{eid}{022301} (\bibinfo{year}{2007}).

\bibitem[{\citenamefont{Lloyd and Braunstein}(1999)}]{braunstein99}
\bibinfo{author}{\bibfnamefont{S.}~\bibnamefont{Lloyd}} \bibnamefont{and}
  \bibinfo{author}{\bibfnamefont{S.~L.} \bibnamefont{Braunstein}},
  \bibinfo{journal}{Phys. Rev. Lett.} \textbf{\bibinfo{volume}{82}},
  \bibinfo{pages}{1784} (\bibinfo{year}{1999}).

\bibitem[{\citenamefont{Bartlett and Sanders}(2002)}]{Bartlett02}
\bibinfo{author}{\bibfnamefont{S.~D.} \bibnamefont{Bartlett}} \bibnamefont{and}
  \bibinfo{author}{\bibfnamefont{B.~C.} \bibnamefont{Sanders}},
  \bibinfo{journal}{Phys. Rev. Lett.} \textbf{\bibinfo{volume}{89}},
  \bibinfo{pages}{207903} (\bibinfo{year}{2002}).

\bibitem[{\citenamefont{Dong et~al.}(2008)\citenamefont{Dong, Lassen, Heersink,
  Marquardt, Filip, Leuchs, and Andersen}}]{dong08}
\bibinfo{author}{\bibfnamefont{R.}~\bibnamefont{Dong}},
  \bibinfo{author}{\bibfnamefont{M.}~\bibnamefont{Lassen}},
  \bibinfo{author}{\bibfnamefont{J.}~\bibnamefont{Heersink}},
  \bibinfo{author}{\bibfnamefont{C.}~\bibnamefont{Marquardt}},
  \bibinfo{author}{\bibfnamefont{R.}~\bibnamefont{Filip}},
  \bibinfo{author}{\bibfnamefont{G.}~\bibnamefont{Leuchs}}, \bibnamefont{and}
  \bibinfo{author}{\bibfnamefont{U.~L.} \bibnamefont{Andersen}},
  \bibinfo{journal}{Nature Physics} \textbf{\bibinfo{volume}{4}},
  \bibinfo{pages}{919} (\bibinfo{year}{2008}).

\bibitem[{\citenamefont{Hage et~al.}(2008)\citenamefont{Hage, Samblowski,
  DiGuglielmo, Franzen, Fiur{\'a}sek, and Schnabel}}]{hage08}
\bibinfo{author}{\bibfnamefont{B.}~\bibnamefont{Hage}},
  \bibinfo{author}{\bibfnamefont{A.}~\bibnamefont{Samblowski}},
  \bibinfo{author}{\bibfnamefont{J.}~\bibnamefont{DiGuglielmo}},
  \bibinfo{author}{\bibfnamefont{A.}~\bibnamefont{Franzen}},
  \bibinfo{author}{\bibfnamefont{J.}~\bibnamefont{Fiur{\'a}sek}},
  \bibnamefont{and} \bibinfo{author}{\bibfnamefont{R.}~\bibnamefont{Schnabel}},
  \bibinfo{journal}{Nature Physics} \textbf{\bibinfo{volume}{4}},
  \bibinfo{pages}{915} (\bibinfo{year}{2008}).

\bibitem[{\citenamefont{Shchukin and Vogel}(2005{\natexlab{a}})}]{shchukin05}
\bibinfo{author}{\bibfnamefont{E.}~\bibnamefont{Shchukin}} \bibnamefont{and}
  \bibinfo{author}{\bibfnamefont{W.}~\bibnamefont{Vogel}},
  \bibinfo{journal}{Phys. Rev. Lett.} \textbf{\bibinfo{volume}{95}},
  \bibinfo{eid}{230502} (\bibinfo{year}{2005}{\natexlab{a}}).

\bibitem[{\citenamefont{Shchukin and Vogel}(2005{\natexlab{b}})}]{shchukin05b}
\bibinfo{author}{\bibfnamefont{E.~V.} \bibnamefont{Shchukin}} \bibnamefont{and}
  \bibinfo{author}{\bibfnamefont{W.}~\bibnamefont{Vogel}},
  \bibinfo{journal}{Phys. Rev. A} \textbf{\bibinfo{volume}{72}},
  \bibinfo{pages}{043808} (\bibinfo{year}{2005}{\natexlab{b}}).

\bibitem[{\citenamefont{Barnett and Phoenix}(1989)}]{barnett89}
\bibinfo{author}{\bibfnamefont{S.~M.} \bibnamefont{Barnett}} \bibnamefont{and}
  \bibinfo{author}{\bibfnamefont{S.}~\bibnamefont{Phoenix}},
  \bibinfo{journal}{Phys. Rev. A} \textbf{\bibinfo{volume}{40}},
  \bibinfo{pages}{2404} (\bibinfo{year}{1989}).

\bibitem[{\citenamefont{Horodecki et~al.}(1996)\citenamefont{Horodecki,
  Horodecki, and Horodecki}}]{horodecki96a}
\bibinfo{author}{\bibfnamefont{R.}~\bibnamefont{Horodecki}},
  \bibinfo{author}{\bibfnamefont{P.}~\bibnamefont{Horodecki}},
  \bibnamefont{and}
  \bibinfo{author}{\bibfnamefont{M.}~\bibnamefont{Horodecki}},
  \bibinfo{journal}{Phys. Lett. A} \textbf{\bibinfo{volume}{210}},
  \bibinfo{pages}{377} (\bibinfo{year}{1996}).

\bibitem[{\citenamefont{Cerf and Adami}(1997)}]{cerf97}
\bibinfo{author}{\bibfnamefont{N.~J.} \bibnamefont{Cerf}} \bibnamefont{and}
  \bibinfo{author}{\bibfnamefont{C.}~\bibnamefont{Adami}},
  \bibinfo{journal}{Phys. Rev. Lett.} \textbf{\bibinfo{volume}{79}},
  \bibinfo{pages}{5194} (\bibinfo{year}{1997}).

\bibitem[{\citenamefont{Vollbrecth and Wolf}(2002)}]{vollbrecht02}
\bibinfo{author}{\bibfnamefont{K.~G.~H.} \bibnamefont{Vollbrecth}}
  \bibnamefont{and} \bibinfo{author}{\bibfnamefont{M.~M.} \bibnamefont{Wolf}},
  \bibinfo{journal}{J. Math. Phys.} \textbf{\bibinfo{volume}{43}},
  \bibinfo{pages}{4299} (\bibinfo{year}{2002}).

\bibitem[{\citenamefont{Giovannetti}(2004)}]{giovannetti04}
\bibinfo{author}{\bibfnamefont{V.}~\bibnamefont{Giovannetti}},
  \bibinfo{journal}{Phys. Rev. A} \textbf{\bibinfo{volume}{70}},
  \bibinfo{pages}{012102} (\bibinfo{year}{2004}).

\bibitem[{\citenamefont{Bialynicki-Birula and Mycielski}(1975)}]{bialynicki75}
\bibinfo{author}{\bibfnamefont{I.}~\bibnamefont{Bialynicki-Birula}}
  \bibnamefont{and}
  \bibinfo{author}{\bibfnamefont{J.}~\bibnamefont{Mycielski}},
  \bibinfo{journal}{Commun. Math. Phys.} \textbf{\bibinfo{volume}{44}},
  \bibinfo{pages}{129} (\bibinfo{year}{1975}).

\bibitem[{\citenamefont{Shannon and Weaver}(1949)}]{shannon}
\bibinfo{author}{\bibfnamefont{C.~E.} \bibnamefont{Shannon}} \bibnamefont{and}
  \bibinfo{author}{\bibfnamefont{W.}~\bibnamefont{Weaver}},
  \emph{\bibinfo{title}{The Mathematical Theory of Communication}}
  (\bibinfo{publisher}{University of Illinois Press}, \bibinfo{year}{1949}).

\bibitem[{\citenamefont{Cover and Thomas}(2006)}]{cover}
\bibinfo{author}{\bibnamefont{Cover}} \bibnamefont{and}
  \bibinfo{author}{\bibnamefont{Thomas}}, \emph{\bibinfo{title}{Elements of
  Information Theory}} (\bibinfo{publisher}{John Wiley and Sons},
  \bibinfo{year}{2006}).

\bibitem[{\citenamefont{Wolf et~al.}(2006)\citenamefont{Wolf, Giedke, and
  Cirac}}]{wolf06}
\bibinfo{author}{\bibfnamefont{M.}~\bibnamefont{Wolf}},
  \bibinfo{author}{\bibfnamefont{G.}~\bibnamefont{Giedke}}, \bibnamefont{and}
  \bibinfo{author}{\bibfnamefont{J.~I.} \bibnamefont{Cirac}},
  \bibinfo{journal}{Phys. Rev. Lett.} \textbf{\bibinfo{volume}{96}},
  \bibinfo{eid}{080502} (\bibinfo{year}{2006}).

\bibitem[{\citenamefont{Rod\'o et~al.}(2008)\citenamefont{Rod\'o, Adesso, and
  Sanpera}}]{rodo08}
\bibinfo{author}{\bibfnamefont{C.}~\bibnamefont{Rod\'o}},
  \bibinfo{author}{\bibfnamefont{G.}~\bibnamefont{Adesso}}, \bibnamefont{and}
  \bibinfo{author}{\bibfnamefont{A.}~\bibnamefont{Sanpera}},
  \bibinfo{journal}{Phys. Rev. Lett.} \textbf{\bibinfo{volume}{100}},
  \bibinfo{eid}{110505} (\bibinfo{year}{2008}).

\bibitem[{\citenamefont{Zyczkowski and Sommers}(2001)}]{zyczkowski01}
\bibinfo{author}{\bibfnamefont{K.}~\bibnamefont{Zyczkowski}} \bibnamefont{and}
  \bibinfo{author}{\bibfnamefont{H.-J.} \bibnamefont{Sommers}},
  \bibinfo{journal}{J. Phys. A: Math and Gen.} \textbf{\bibinfo{volume}{34}},
  \bibinfo{pages}{7111} (\bibinfo{year}{2001}).

\end{thebibliography}

%%%%%%%%%%%%%%%%%%%%%%%%%%%%%%%%%%%%%%%%%%%%%%%%%%%%%%%%%%%%%%%%%%

\end{document}